\documentclass[twocolumn,showpacs,preprintnumbers,amsmath,amssymb,showkeys]{revtex4}

\usepackage{graphicx}
\usepackage{dcolumn}
\usepackage{bm}

\begin{document} 
 
\noindent

\preprint{}

\title{A stochastic model for quantum measurement}

\author{Agung Budiyono}
\email{agungbymlati@gmail.com}

\affiliation{Jalan Emas 772 Growong Lor RT 04 RW 02 Juwana, Pati, 59185 Jawa Tengah, Indonesia}
 
\date{\today}
  
\begin{abstract}    
We develop a statistical model of microscopic stochastic deviation from classical mechanics based on a stochastic processes with a transition probability that is assumed to be given by an exponential distribution of infinitesimal stationary action. We apply the statistical model to stochastically modify a classical mechanical model for the measurement of physical quantities reproducing the prediction of quantum mechanics. The system+apparatus always have a definite configuration all the time as in classical mechanics, fluctuating randomly following a continuous trajectory. On the other hand, the wave function and quantum mechanical Hermitian operator corresponding to the physical quantity arise formally as artificial mathematical constructs. During a single measurement, the wave function of the whole system+apparatus evolves according to a Schr\"odinger equation and the configuration of the apparatus acts as the pointer of the measurement so that there is no wave function collapse. We will also show that while the result of each single measurement event does not reveal the actual value of the physical quantity prior to measurement, its average in an ensemble of identical measurement is equal to the average of the actual value of the physical quantity prior to measurement over the distribution of the configuration of the system.     
\end{abstract} 
  
\maketitle   

\section{Motivation}   
 
Despite of the superb empirical successes of quantum mechanics with a broad of practical applications in the development of technology, there are still many contradicting views on the meaning of the theory \cite{Isham book}. Such an absence of consensus on the conceptual foundation of quantum mechanics, after almost nine decades since its completion, might be due to the fact that the numerous postulates of standard quantum mechanics are highly `abstract and formal-mathematical' with non-transparent physical and operational meaning. First, standard quantum mechanics does not provide a transparent explanation on the status of the wave function with regard to the state of the system under interest. Is the wave function physical or merely an artificial mathematical construct? Nor it gives a transparent physical and/or operational explanation why the physical observable quantities are represented by certain Hermitian operators, and which part of experiment that corresponds to their measurement. Equally importantly, one may also ask: when is a given Hermitian operator a representation of a meaningful or observable physical quantity? The above couple of problems are intimately related to the problem of the meaning and origin of the abstract and ``strange'' \cite{Giulini strange} quantization procedures: canonical quantization, path integral, etc., via which quantum systems can be obtained from the corresponding classical systems.  
 
Further, the postulates of quantum mechanics give an {\it algorithm} to accurately predict/calculate the statistical results in an ensemble of identical measurement, rather then describes what is really happening physically and operationally in each single measurement event. This naturally raises several conceptual problems. First, one could ask: why measurement is dictated by a random, discontinuous and non-unitary time evolution, that of the postulate of wave function collapse, while when unobserved, a closed system follows a deterministic, continuous and unitary time evolution given by the Schr\"odinger equation. Providing a unified law for the time evolution of the system, observed or not, is the main issue of the infamous `measurement problem'. The algorithm also does not give us a clue on the status or meaning of each single measurement result, and the statistics of the results in an ensemble of identical measurement, with regard to the properties of the system prior to measurement. Nor, the postulates tell us {\it when} and {\it where} the random collapse of wave function occurs and {\it what} is needed for its occurrence (what constitutes a measurement?) It is also physically not clear {\it why} the statistical results follow the Born's rule. Moreover, to be precise, the apparatus is also composed of elementary particles so that the whole system+apparatus should be regarded as a legitimate closed quantum system which must evolve according to a Schr\"odinger equation. The linearity of the latter however allows for the superposition of macroscopically distinguishable states of the pointer of the apparatus which, assuming that the wave function is physical and complete, leads to `the paradox of Schr\"odinger's cat'.          
 
To solve the above foundational problems, there is a growing interest recently in the program to reconstruct quantum mechanics. In this program, rather than directly pursuing interpretational questions on the abstract mathematical structures of quantum mechanics, one asks along with Wheeler: ``why the quantum?'' \cite{Wheeler}, and wonders if the numerous abstract postulates of quantum mechanics can be derived from a concrete and transparent physical model. It is also of great interest if such a physical model can be further derived uniquely from a set of conceptually simple and physically transparent axioms \cite{Shimony 2,Popescu no-signaling 1,Grunhaus no-signaling,Popescu no-signaling 2}. A lot of progresses along this line of approach has been made either within realist \cite{Nelson,Garbaczewski,Markopoulou,dela Pena 1} or operational/information theoretical frameworks \cite{Rovelli,Zeilinger information quantization,Hardy construction,Simon,Clifton,Pawlowski,Navascues,Brukner2,Masanes,Chiribella,Torre,Fivel}. In the former, quantum fluctuations is assumed to be physically real and objective, and thus should be properly modeled by some stochastic processes. On the other hand, in the latter, quantum fluctuations is assumed to be fundamentally related to the concept of information and its processing. One then  searches for a set of basic features of information processing which can be promoted as axioms to reconstruct quantum mechanics. One of the advantages of the reconstruction of quantum mechanics is that it might give useful physical insight to extend quantum mechanics either by changing the axioms or varying the free parameters of the physical model.     
 
In the present paper, we shall develop a statistical model of stochastic deviation from classical mechanics in microscopic regime based on a Markovian stochastic processes to reconstruct quantum mechanics. This is done by assuming that the transition probability between two infinitesimally close spacetime points in configuration space via a random path is given by an exponential distribution of infinitesimal stationary action. We then apply the statistical model to stochastically modify a classical mechanical model of measurement so that the whole `system+apparatus' is subjected to the stochastic fluctuations of infinitesimal stationary action. We shall show, by giving an explicit example of the measurement of angular momentum, that the model reproduces the prediction of quantum mechanics. 

Unlike canonical quantization, the system possesses a definite configuration all the time as in classical mechanics, following a continuous trajectory randomly fluctuating with time. The configuration of the system should thus be regarded as the beable  of the theory in Bell's sense \cite{Bell beable}. The Hermitian differential operator corresponding to the angular momentum and the wave function, on the other hand, arise formally and simultaneously as artificial convenient mathematical tools as one works in the Hilbert space representation. During a single measurement event, the wave function of the whole system+apparatus follows a Schr\"odinger equation, and as in classical mechanics, the configuration of part of the apparatus plays the role as pointer so that there is no wave function collapse. We will also show that while the result of each single measurement event does not reveal the actual value of the angular momentum prior to measurement, its average in an ensemble of identical measurement is equal to the average of the actual value of the angular momentum prior to measurement over the distribution of the configuration of the system.

Without giving the technical detail, we shall also argue that the same conclusion carries over the measurement of position and angular momentum. In this paper, we shall confine the discussion to system of spin-less particles.  
  
\section{A statistical model of microscopic stochastic deviation from classical mechanics}   

\subsection{A class of stochastic processes in microscopic regime based on random fluctuations of infinitesimal stationary action}

There is a wealth of evidences that in microscopic regime the deterministic classical mechanics suffers a stochastic correction. Yet, the prediction of quantum mechanics on the AB (Aharonov-Bohm) effect \cite{Aharonov-Bohm} and its experimental verification \cite{Peshkin} suggest that the randomness in microscopic regime is inexplicable in term of conventional random forces as in the Brownian motion. To describe such a microscopic randomness, let us develop the following stochastic model. Let us consider a system of particles whose configuration is denoted by $q$ and its evolution is parameterized by time $t$. Let us assume that the Lagrangian depends on a randomly fluctuating variable $\xi$: $L=L(q,\dot{q};\xi)$, whose origin is not our present concern. Let us assume that the time scale for the fluctuations of $\xi$ is $dt$.

Let us then consider two infinitesimally close spacetime points $(q;t)$ and $(q+dq;t+dt)$ such that $\xi$ is constant. Let us assume that fixing $\xi$, the principle of stationary action is valid to select a segment of path, denoted by $\mathcal{J}(\xi)$,  that connects the two points. One must then solve a variational problem with fixed end points: $\delta(Ldt)=0$. This variational problem leads to the existence of a function $A(q;t,\xi)$, the Hamilton principle function, whose differential along the path is given by \cite{Rund book}, for a fixed $\xi$, 
\begin{equation} 
dA=Ldt=p\cdot dq-Hdt, 
\label{infinitesimal stationary action} 
\end{equation} 
where $p(\dot{q})=\partial L/\partial{\dot{q}}$ is the momentum and $H(q,p)\doteq p\cdot\dot{q}(p)-L(q,\dot{q}(p))$ is the Hamiltonian. Here we have made an assumption that the Lagrangian is not singular $\mbox{det}(\partial^2 L/\partial\dot{q}_i\partial\dot{q}_j)\neq 0$. The above relation implies the following Hamilton-Jacobi equation:
\begin{equation}
p=\partial_qA\hspace{2mm}\&\hspace{2mm}-H(q,p)=\partial_tA, 
\label{Hamilton-Jacobi condition}
\end{equation}
which is valid during a microscopic time interval in which $\xi$ is fixed. Hence, $dA(\xi)$ is just the `infinitesimal stationary action' along the corresponding short path during the infinitesimal time interval $dt$ in which $\xi$ is fixed. 

Varying the value of $\xi$, the principle of stationary action will therefore pick up various different paths $\mathcal{J}(\xi)$, all connecting the same two infinitesimally close spacetime points in configuration space, each with different values of infinitesimal stationary action $dA(\xi)$. $dA(\xi)$ thus is randomly fluctuating due to the fluctuations of $\xi$. Hence, we have a stochastic processes in which the system starting with configuration $q$ at time $t$ may take various different paths randomly to arrive at $q+dq$ at time $t+dt$. Assuming that the stochastic processes in Markovian, then it is completely determined by a `transition probability' for the system starting with configuration $q$ at time $t$ to move to its infinitesimally close neighbor $q+dq$ at time $t+dt$ via a path $\mathcal{J}(\xi)$, below denoted by 
\begin{equation}
P((q+dq;t+dt)|\{\mathcal{J}(\xi),(q;t)\}).  
\label{transition probability}
\end{equation}
Since the stochastic processes is supposed to model a stochastic deviation from classical mechanics, then it is reasonable to assume that the transition probability is a function of a quantity that measures the deviation from classical mechanics. 
 
It is natural to express the transition probability in term of the stochastic quantity $dA(\xi)$ evaluated along the short segment of trajectory. To do this, let us first assume that $\xi$ is the simplest random variable with two possible values, a binary random variable. Without losing generality let us assume that the two possible values of $\xi$ differ from each other only by their signs, namely one is the opposite of the other, $\xi=\pm|\xi|$. Suppose that both realizations of $\xi$ lead to the same path so that $dA(\xi)=dA(-\xi)$. Since the stationary action principle is valid for both values of $\pm\xi$, then such a model must recover classical mechanics. Hence, the non-classical behavior should correspond to the case when the different signs of $\xi$ lead to different trajectories so that $dA(\xi)\neq dA(-\xi)$. 

Now let us proceed to assume that $\xi$ may take continuous values. Let us assume that even in this case the magnitude of the difference of the value of $dA$ at $\pm\xi$, 
\begin{equation}
Z(q;t,\xi)\doteq dA(q;t,\xi)-dA(q;t,-\xi)=-Z(q;t,-\xi),  
\end{equation}
measures the non-classical behavior of the stochastic process, namely the larger the difference, the stronger is the deviation from classical mechanics. Hence $Z(\xi)$ is randomly fluctuating due to the fluctuations of $\xi$, and we shall use its distribution as the transition probability that we are looking for to construct the stochastic model: 
\begin{equation}
P((q+dq;t+dt)|\{\mathcal{J}(\xi),(q;t)\})=P(Z(\xi)). 
\end{equation}

It is evident that the randomness is built into the statistical model in a fundamentally different way from that of the classical Brownian motion. Unlike the latter in which the randomness is introduced by adding some random forces, the model is based on a stochastic fluctuations of infinitesimal stationary action. Hence, we have assumed that the Lagrangian formalism based on energies is more fundamental than Newtonian formalism based on forces. One may expect that this will explain the physical origin of AB effect.

\subsection{Exponential distribution of infinitesimal stationary action as the transition probability}

How is then $Z(\xi)$ distributed? First, it is reasonable to assume that the transition probability must be decreasing as the non-classicality becomes stronger. Hence, the transition probability must be a decreasing function of the absolute value of $Z(\xi)$. There are infinitely many such probability distribution functions. Below for the reason that will become clear later, we shall assume that $Z(\xi)$ is distributed according to the following exponential law:
\begin{eqnarray}
P(Z)\propto N e^{\frac{1}{\lambda(\xi)}Z(\xi)}=N e^{\frac{1}{\lambda(\xi)}\big(dA(\xi)-dA(-\xi)\big)},
\label{exponential law 1st}
\end{eqnarray}
where $N$ is a factor independent of $Z(\xi)$ whose form to be specified later and $\lambda(\xi)$ is a non-vanishing function of $\xi$ with action dimensional, thus is randomly fluctuating. Note that by definition, $Z(\xi)$ changes its sign as $\xi$ flips its sign: $Z(-\xi)=-Z(\xi)$. On the other hand, to guarantee the negative definiteness of the exponent in Eq. (\ref{exponential law 1st}) for normalizability, $\lambda$ must always have the opposite sign of $Z(\xi)$. This demands that $\lambda$ must flip its sign as $\xi$ changes its sign. This fact therefore allows us to assume that both $\lambda(\xi)$ and $\xi$ always have the {\it same} sign. Hence the time scale for the fluctuations of the sign of $\lambda$ must be the same as that of $\xi$ given by $dt$. 

However, it is clear that for the distribution of Eq. (\ref{exponential law 1st}) to make sense mathematically, the time scale for the fluctuations of $|\lambda|$, denoted by $\tau_{\lambda}$, must be much larger than that of $|\xi|$, $\tau_{\xi}$. Let us further assume that $\tau_{\xi}$ is much larger than $dt$. One thus has 
\begin{equation}
\tau_{\lambda}\gg\tau_{\xi}\gg dt. 
\label{time scales}
\end{equation}
Hence, in a time interval of length $\tau_{\xi}$, the absolute value of $\xi$ can be regarded constant while its sign may fluctuate randomly together with the sign of $\lambda$ in a time scale given by $dt$. Moreover, in a time interval $\tau_{\lambda}$, $|\lambda|$ is constant and $|\xi|$ fluctuates randomly so that the distribution of $Z(\xi)$ is given by the exponential law of Eq. (\ref{exponential law 1st}) characterized by $|\lambda|$.    

Next, let us introduce a new stochastic quantity $S(q;t,\xi)$ so that the differential along the path $\mathcal{J}(\xi)$ is given by 
\begin{equation}
dS(q;t,\xi)=\frac{dA(q;t,\xi)+dA(q;t,-\xi)}{2}=dS(q;t,-\xi).  
\label{infinitesimal symmetry}
\end{equation}
Subtracting $dA(q;t,\xi)$ to both sides, one has 
\begin{equation}
dS(q;t,\xi)-dA(q;t,\xi)=\frac{dA(q;t,-\xi)-dA(q;t,\xi)}{2}.
\label{average of classical deviation}
\end{equation}  
Using $dS$, the transition probability of Eq. (\ref{exponential law 1st}) can thus be written as 
\begin{eqnarray} 
P((q+dq;t+dt)|\{\mathcal{J}(\xi),(q;t)\})\hspace{15mm}\nonumber\\
\propto N e^{-\frac{2}{\lambda}(dS(q;t,\xi)-dA(q;t,\xi))}\doteq P_S(dS|dA).  
\label{exponential distribution of DISA}
\end{eqnarray}  
 
Since $dA(\xi)$ is just the infinitesimal stationary action along the path $\mathcal{J}(\xi)$, then we shall refer to $dS(\xi)-dA(\xi)$ as a deviation from infinitesimal stationary action. One may therefore see the above transition probability to be given by an exponential distribution of deviation from infinitesimal stationary action $dS-dA$ parameterized by $|\lambda|$. It can also be regarded as the conditional probability density of $dS$ given $dA$, suggesting the use of the notation $P_S(dS|dA)$. The relevancy of such an exponential law to model microscopic stochastic deviation from classical mechanics is firstly suggested in Ref. \cite{AgungSMQ4}.  

Further, there is no a priori reason on how the sign of $dS(\xi)-dA(\xi)$, which is equal to the sign of $Z(-\xi)=dA(-\xi)-dA(\xi)$ due to Eq. (\ref{average of classical deviation}), should be distributed at any given spacetime point. The principle of insufficient reason (principle of indifference) \cite{Jaynes book} then suggests to assume that at any spacetime point, there is equal probability for $dS-dA$ to take positive or negative values. Since the sign of $dS(\xi)-dA(\xi)$ changes as $\xi$ flips its sign, then the sign of $\xi$ must also be distributed equally probably. Hence, the probability density of the occurrence of $\xi$ at any time, denoted below by $P_{H}(\xi)$, must satisfy the following unbiased condition:  
\begin{equation}
P_{H}(\xi)=P_{H}(-\xi).  
\label{God's unbiased}
\end{equation}  
Let us note that $P_{H}(\xi)$ may depend on time, thus it is in general not stationary. Since the sign of $\lambda$ is always the same as that of $\xi$, then the probability for the occurrence of $\lambda$ must also satisfy the same unbiased condition
\begin{equation}
P(\lambda)=P(-\lambda).  
\end{equation}
Moreover, from Eq. (\ref{infinitesimal symmetry}), one obtains, for a fixed value of $\xi$, the following symmetry relations: 
\begin{eqnarray}
\partial_qS(q;t,\xi)=\partial_qS(q;t,-\xi),\nonumber\\
\partial_tS(q;t,\xi)=\partial_tS(q;t,-\xi). 
\label{quantum phase symmetry}
\end{eqnarray} 

Fixing $|\lambda|$ in Eq. (\ref{exponential distribution of DISA}), then the average deviation from infinitesimal stationary action is given by
\begin{equation}
\overline{dS-dA}=|\lambda|/2. 
\label{the P} 
\end{equation}
One can then see that in the regime where the average deviation from infinitesimal stationary action is much smaller than the infinitesimal stationary action itself, namely $|dA(\xi)|/|\lambda|\gg 1$, or formally in the limit $|\lambda|\rightarrow 0$, Eq. (\ref{exponential distribution of DISA}) reduces to  
\begin{equation}
P_S(dS|dA)\rightarrow \delta(dS-dA), 
\label{macroscopic classicality}
\end{equation}
or $dS(\xi)\rightarrow dA(\xi)$, so that $S$ satisfies the Hamilton-Jacobi equation of (\ref{Hamilton-Jacobi condition}) by the virtue of Eq. (\ref{infinitesimal stationary action}). Due to Eq. (\ref{average of classical deviation}), in this regime one also has $dA(\xi)=dA(-\xi)$. Hence such a limiting case must be identified to correspond to the macroscopic regime. This further suggests that $|\lambda|$ must take a microscopic value.   

\subsection{Stochastic modification of Hamilton-Jacobi equation}
 
Let us now derive a set of differential equations which characterizes the stochastic processes when the transition probability is given by Eq. (\ref{exponential distribution of DISA}). Let us consider a time interval of length $\tau_{\lambda}$ in which $|\lambda|$ is effectively constant. Recall that since $\tau_{\lambda}\gg\tau_{\xi}\gg dt$, then within this time interval, $dS(\xi)-dA(\xi)$ fluctuates randomly due to the fluctuations of $\xi$, distributed according to the exponential law of Eq. (\ref{exponential distribution of DISA}) characterized by $|\lambda|$. 

Let us then denote the joint-probability density that at time $t$ the configuration of the system is $q$ and a random value of $\xi$ is realized by $\Omega(q,\xi;t)$. The marginal probability densities are thus given by 
\begin{eqnarray}
\rho(q;t)\doteq\int d\xi\Omega(q,\xi;t),\hspace{2mm}P_{H}(\xi)=\int dq \Omega(q,\xi;t).\hspace{0mm}
\label{marginal probabilities general}
\end{eqnarray} 
To comply with Eq. (\ref{God's unbiased}), the joint-probability density must satisfy the following symmetry relation: 
\begin{eqnarray} 
\Omega(q,\xi;t)=\Omega(q,-\xi;t). 
\label{God's fairness}
\end{eqnarray} 
Both Eqs. (\ref{God's fairness}) and (\ref{quantum phase symmetry}) will play important roles later. 

Let us then evolve $\Omega(q,\xi;t)$ along a time interval $\Delta t$ with $\tau_{\xi}\ge\Delta t\gg dt$ so that the absolute value of $\xi$ is effectively constant while its sign may fluctuate randomly. Given a fixed value of $\xi$, let us consider two infinitesimally close spacetime points $(q;t)$ and $(q+dq;t+dt)$. Let us assume that for this value of $\xi$, the two points are connected to each other by a segment of trajectory $\mathcal{J}(\xi)$ picked up by the principle of stationary action so that the differential of $S(\xi)$ along this segment is $dS(\xi)$, parameterized by $\xi$. Then for a fixed value of $\xi$, according to the conventional probability theory, the conditional joint-probability density that the system initially at $(q;t)$ traces the segment of trajectory $\mathcal{J}(\xi)$ and end up at $(q+dq;t+dt)$, denoted below as $\Omega\big(\{(q+dq,\xi;t+dt),(q,\xi;t)\}\big|\mathcal{J}(\xi)\big)$, is equal to the probability that the configuration of the system is $q$ at time $t$, $\Omega(q,\xi;t)$, multiplied by the transition probability between the two infinitesimally close points via the segment of trajectory $\mathcal{J}(\xi)$ which is given by Eq. (\ref{exponential distribution of DISA}). One thus has  
\begin{eqnarray}
\Omega\Big(\{(q+dq,\xi;t+dt),(q,\xi;t)\}\big|\mathcal{J}(\xi)\Big)\hspace{15mm}\nonumber\\
=P((q+dq;t+dt)|\{\mathcal{J}(\xi),(q;t)\})\times\Omega(q,\xi;t)\nonumber\\
\propto Ne^{-\frac{2}{\lambda(t)}(dS(\xi)-dA(\xi))}\times\Omega(q,\xi;t).  
\label{probability density} 
\end{eqnarray}    

The above equation describing the dynamics of ensemble of trajectories must give back the time evolution of classical mechanical ensemble of trajectories when $S$ approaches $A$. This requirement puts a constraint on the functional form of the factor $N$ in Eq. (\ref{exponential distribution of DISA}). To see this, let us assume that $N$ takes the following general form: 
\begin{equation}
N\propto\exp(-\theta(S)dt), 
\label{exponential classical}
\end{equation}
where $\theta$ is a scalar function of $S$. Inserting this into Eq. (\ref{probability density}), taking the limit $S\rightarrow A$ and expanding the exponential up to the first order one gets $\Omega\big(\{(q+dq,\xi;t+dt),(q,\xi;t)\}\big|\mathcal{J}(\xi)\big)\approx \big[1-\theta(A)dt\big]\Omega(q,\xi;t)$, which can be further written as 
\begin{eqnarray}
d\Omega=-\big(\theta(A)dt\big)\Omega, 
\label{fundamental equation 0}
\end{eqnarray} 
where $d\Omega(q,\xi;t)\doteq\Omega\big(\{(q+dq,\xi;t+dt),(q,\xi;t)\}\big|\mathcal{J}(\xi)\big)-\Omega(q,\xi;t)$ is the change of the probability density $\Omega$ due to the transport along the segment of trajectory $\mathcal{J}(\xi)$. Dividing both sides by $dt$ and taking the limit $dt\rightarrow 0$, one obtains $\dot{\Omega}+\theta(A)\Omega=0$. To guarantee a smooth correspondence with classical mechanics, the above equation must be identified as the continuity equation describing the dynamics of ensemble of classical trajectories. To do this, it is sufficient to choose $\theta(S)$ to be determined uniquely by the classical Hamiltonian as \cite{AgungSMQ4}  
\begin{equation}
\theta(S)=\partial_q\cdot\Big(\frac{\partial H}{\partial p}\Big|_{p=\partial_qS}\Big), 
\label{QC correspondence}
\end{equation}
so that in the limit $S\rightarrow A$, it is given by the divergence of a classical velocity field.  

Now, let us consider the case when $|(dS-dA)/\lambda|\ll 1$. Again, inserting Eq. (\ref{exponential classical}) into Eq. (\ref{probability density}) and expanding the exponential on the right hand side up to the first order one gets
\begin{eqnarray}
d\Omega=-\Big[\frac{2}{\lambda}(d S-dA)+\theta(S)d t\Big]\Omega.    
\label{fundamental equation 0}
\end{eqnarray} 
Further, recalling that $\xi$ is fixed during the infinitesimal time interval $dt$, one can expand the differentials $d\Omega$ and $dS$ in Eq. (\ref{fundamental equation 0}) as $dF=\partial_tF dt+\partial_qF\cdot dq$. Using Eq. (\ref{infinitesimal stationary action}), one finally obtains the following pair of coupled differential equations:  
\begin{eqnarray}
p(\dot{q})=\partial_qS+\frac{\lambda}{2}\frac{\partial_q\Omega}{\Omega},\hspace{8mm}\nonumber\\
-H(q,p)=\partial_tS+\frac{\lambda}{2}\frac{\partial_t\Omega}{\Omega}+\frac{\lambda}{2}\theta(S). 
\label{fundamental equation rederived}
\end{eqnarray} 

Some notes are in order. First, the above pair of relations are valid when $\xi$ is fixed. However, since as discussed above, $P_S(dS|dA)$ is insensitive to the sign of $\xi$ which is always equal to the sign of $\lambda$, then the above pair of equations are valid in a microscopic time interval of length $\tau_{\xi}$ during which the magnitude of $\xi$, and also $\lambda$ due to Eq. (\ref{time scales}), are constant while their signs may change randomly. To have an evolution for a finite time interval $\tau_{\lambda}>t>\tau_{\xi}$, one can proceed along the following approximation. First, one divides the time into a series of intervals of length $\tau_{\xi}$: $t\in[(k-1)\tau_{\xi},k\tau_{\xi})$, $k=1,2,\dots$, and attributes to each interval a random value of $\xi(t)=\xi_k$ according to the probability distribution $P_{H_k}(\xi_k)=P_{H_k}(-\xi_k)$. Hence, during the interval $[(k-1)\tau_{\xi},k\tau_{\xi})$, the magnitude of $\xi=\xi_k$ is kept constant while its sign may change randomly in a time scale $dt$, so that Eq. (\ref{fundamental equation rederived}) is valid. One then applies the pair of equations in (\ref{fundamental equation rederived}) during each interval of time with fixed $|\xi(t)|=|\xi_k|$, consecutively. Moreover, to have a time evolution for $t\ge \tau_{\lambda}$, one must now take into account the fluctuations of $|\lambda|$ with time.      

It is evident that as expected, in the formal limit $\lambda\rightarrow 0$, Eq. (\ref{fundamental equation rederived}) reduces back to the Hamilton-Jacobi equation of (\ref{Hamilton-Jacobi condition}). In this sense, Eq. (\ref{fundamental equation rederived}) can be regarded as a generalization of the Hamilton-Jacobi equation due to the stochastic deviation from infinitesimal stationary action following the exponential law of Eq. (\ref{exponential distribution of DISA}). Unlike the Hamilton-Jacobi equation in which we have a single unknown function $A$, however, to calculate the velocity or momentum and energy, one now needs a pair of unknown functions $S$ and $\Omega$. The relations in Eq. (\ref{fundamental equation rederived}) must not be interpreted that the momentum and energy of the particles are determined causally by the gradient of the probability density $\Omega$ (or $\ln(\Omega)$), which is physically absurd. Rather it is the other way around as shown explicitly by Eq. (\ref{fundamental equation 0}). The relation is thus kinematical rather than causal-dynamical.         

Let us then consider a system of two non-interacting particles whose configuration is denoted by $q_1$ and $q_2$. The Lagrangian is thus decomposable as $L(q_1,q_2,\dot{q}_1,\dot{q}_2)=L_1(q_1,\dot{q}_1)+L_2(q_2,\dot{q}_2)$, so that the infinitesimal stationary action is also decomposable: $dA(q_1,q_2)=dA_1(q_1)+dA_2(q_2)$, and accordingly one has $dS(q_1,q_2)=dS_1(q_1)+dS_2(q_2)$ by the virtue of Eq. (\ref{infinitesimal symmetry}). On the other hand, since the classical Hamilton $H$ is decomposable as $H(q_1,q_2,p_1,p_2)=H_1(q_1,p_1)+H_2(q_2,p_2)$, $p_{i}$, $i=1,2$, is the classical momentum of the $i-$particle, then $\theta$ of Eq. (\ref{QC correspondence}) is also decomposable: $\theta(q_1,q_2)=\theta_1(q_1)+\theta_2(q_2)$. Inserting all these into Eqs. (\ref{exponential classical}) and (\ref{exponential distribution of DISA}), then one can see that the distribution of deviation from infinitesimal stationary action for the non-interacting two particles is separable as 
\begin{equation}
P_S(dS_1+dS_2|dA_1+dA_2)=P_S(dS_1|dA_1)P_S(dS_2|dA_2).  
\label{non-interacting vs separability}
\end{equation}
Namely the joint-probability distribution of the deviations from infinitesimal stationary action of the two particles system is separable into the probability distribution of the deviation with respect to each single particle. They are thus independent of each other, as intuitively expected for non-interacting particles. It is interesting to remark that the above statistical separability for non-interacting particles is unique to the exponential law. A Gaussian distribution of deviation from infinitesimal stationary action for example does not have such a property. 
 
\section{Stochastic processes for quantum measurement}

In this section, we shall apply the above statistical model of microscopic stochastic deviation from classical mechanics to a classical mechanical model of measurement consisting of two interacting particles, one is regarded as the system whose physical properties are being measured and the other plays the role as the apparatus pointer. We have to admit that such a model of measurement apparatus by a single particle is too simple: it is inspired by rather than describes completely a realistic measurement. Nevertheless, we shall show that the model reproduces the prediction of standard quantum mechanics. We also believe that the approach can in principle be generalized to model realistic macroscopic apparatus. See for example Ref. \cite{Theo} for richer models of quantum measurement with a realistic apparatus.   

In the paper we shall only give the details for the measurement of angular momentum. The measurement of position and linear momentum can be done in exactly the same way. 

For completeness of the presentation, in the appendix, we reproduce the application of the statistical model to a system of particles subjected to potentials with a Hamiltonian that is quadratic in momentum reported in Ref. \cite{AgungSMQ4}. Several subtle issues in Ref. (\cite{AgungSMQ4}) are clarified.    

\subsection{Measurement in classical mechanics}

Let us first briefly discuss the essential points of a measurement model in classical mechanics consisting of two interacting particles. To do this, let us assume that the interaction classical Hamiltonian is given by 
\begin{equation}
H_I=gO_1(q_1,p_1){p}_2. 
\label{classical Hamiltonian measurement}
\end{equation}
Here $g$ is an interaction coupling and $O_1(q_1,p_1)$ is a physical quantity referring to the first particle. Let us further assume that the interaction is impulsive ($g$ is sufficiently strong) so that the single particle Hamiltonians of each particle are ignorable. The discussion on single particle Hamiltonian is given in the appendix. 
 
The interaction Hamiltonian above can be used as a classical mechanical model of measurement of the classical physical quantity $O_1(q_1,p_1)$ of the first particle by regarding the position of the second particle as the pointer of the apparatus of measurement. To see this, first, in such a model $O_1$ is conserved: $\dot{O}_1=\{O_1,H_I\}=0$ where $\{\cdot,\cdot\}$ is the usual Poisson bracket. The interaction Hamiltonian of Eq. (\ref{classical Hamiltonian measurement}) then correlates the value of $O_1$ with the momentum of the apparatus ${p}_2$ while keeping the value of $O_1$ unchanged. On the other hand, one also has $\dot{q}_2=\{q_2,H_I\}=gO_1$, which, by the virtue of the fact that $O_1$ is a constant of motion, can be integrated to give 
\begin{equation}
q_2(t_M)=q_2(0)+gO_1t_M, 
\label{classical-quantum pointer}
\end{equation}
where $t_M$ is the time span of the measurement-interaction. The value of $O_1$ prior to the measurement can thus in principle be inferred from the observation of the initial and final values of $q_2$. 
 
In this way, the measurement of the physical quantity $O_1(q_1,p_1)$ of the first particle is reduced to the measurement of the position of the second particle $q_2$. In the model, $q_2(t)$ therefore plays the role as the pointer of the apparatus of measurement. This is in principle what is actually done in experiment, either involving macroscopic or microscopic objects, where one reads the position of the needle in the meter or the position of the detector that `clicks', etc. It is thus assumed that measurement of position can in principle be done straightforwardly. To have a physically and operationally smooth quantum-classical correspondence, we shall keep this `operationally clear' measurement mechanism while we proceed below to subject the classical system to a stochastic fluctuations of infinitesimal stationary action according to the statistical model discussed in the previous section. 

It is also obvious from the above exposition that in classical mechanics, each single measurement event reveals the value of the physical quantity under interest prior to the measurement up to the precision of position measurement of the pointer. In particular, there is a one to one mapping between the continuous values of the pointer $q_2(t_M)$ and the continuous possible values of the physical quantity being measured $O_1(q_1,p_1)$ prior to the measurement. We shall show below that in the statistical model, this is in general no more the case. 

\subsection{The Schr\"odinger equation in measurement of angular momentum}
 
Now let us apply the statistical model discussed in the previous section to stochastically modify the above classical mechanical model of measurement. Let us first consider a time interval of length $\tau_{\lambda}$ in which the absolute value of  $\lambda$ is effectively constant while its sign is allowed to fluctuate randomly together with the random fluctuations of the sign of $\xi$. Let us then divide it into a series of microscopic time intervals of length $\tau_{\xi}$, $[(k-1)\tau_{\xi},k\tau_{\xi})$, $k=1,2,\dots$ and attribute to each interval a random value of $\xi(t)=\xi_k$ according to the probability distribution $P_{H_k}(\xi_k)=P_{H_k}(-\xi_k)$ so that in each interval, the magnitude of $\xi$ is constant while its sign is allowed to change randomly. During each time interval $[(k-1)\tau_{\xi},k\tau_{\xi})$, the pair of equations in Eq. (\ref{fundamental equation rederived}), each with constant value of $|\xi_k|$, thus apply. 

For concreteness, let us consider the measurement of the $z-$part angular momentum of the first particle. The classical interaction Hamiltonian of Eq. (\ref{classical Hamiltonian measurement}) then reads      
\begin{equation}
H_I=gl_{z_1}{p}_2,\hspace{2mm}\mbox{with}\hspace{2mm}l_{z_1}=x_1{p}_{y_1}-y_1{p}_{x_1}.  
\label{classical Hamiltonian measurement angular momentum}
\end{equation} 
Let us first consider a microscopic time interval $[(k-1)\tau_{\xi},k\tau_{\xi})$. Using the above form of $H_I$ to express $\dot{q}$ in term of $p$ via the (kinematic part of the) usual Hamilton equation $\dot{q}=\partial H/\partial p$, the upper equation of (\ref{fundamental equation rederived}) becomes 
\begin{eqnarray}
\dot{x}_1=-gy_1\Big(\partial_{q_2}S+\frac{\lambda}{2}\frac{\partial_{q_2}\Omega}{\Omega}\Big),\hspace{2mm}\dot{y}_1=gx_1\Big(\partial_{q_2}S+\frac{\lambda}{2}\frac{\partial_{q_2}\Omega}{\Omega}\Big),\hspace{2mm}\nonumber\\
\dot{q}_2=g\Big(x_1\Big(\partial_{y_1}S+\frac{\lambda}{2}\frac{\partial_{y_1}\Omega}{\Omega}\Big)-y_1\Big(\partial_{x_1}S+\frac{\lambda}{2}\frac{\partial_{x_1}\Omega}{\Omega}\Big)\Big),\hspace{5mm}\nonumber\\
\label{velocity angular momentum}
\end{eqnarray} 
and $\dot{z}_1=0$. Assuming that the probability is conserved, one gets, after a simple calculation, the following continuity equation: 
\begin{eqnarray}
0=\partial_t\Omega+\partial_q\cdot(\dot{q}\Omega)\hspace{60mm}\nonumber\\
=\partial_t\Omega-gy_1\partial_{x_1}(\Omega\partial_{q_2}S)+gx_1\partial_{y_1}(\Omega\partial_{q_2}S)+gx_1\partial_{q_2}(\Omega\partial_{y_1}S)\nonumber\\
-gy_1\partial_{q_2}(\Omega\partial_{x_1}S)-g\lambda(y_1\partial_{x_1}\partial_{q_2}\Omega-x_1\partial_{y_1}\partial_{q_2}\Omega).\nonumber\\
\label{FPE angular momentum}
\end{eqnarray}

On the other hand, from Eq. (\ref{classical Hamiltonian measurement angular momentum}), $\theta(S)$ of Eq. (\ref{QC correspondence}) is given by  
\begin{equation}
\theta(S)=2g(x_1\partial_{q_2}\partial_{y_1}S-y_1\partial_{q_2}\partial_{x_1}S).
\end{equation}
Substituting this into the lower equation of (\ref{fundamental equation rederived}), one then obtains 
\begin{eqnarray}
-H_I(q,p(\dot{q}))=\partial_{t}S+\frac{\lambda}{2}\frac{\partial_{t}\Omega}{\Omega}\nonumber\\
+ g\lambda(x_1\partial_{y_1}\partial_{q_2}S-y_1\partial_{x_1}\partial_{q_2}S).\hspace{0mm}
\label{fundamental equation angular momentum}
\end{eqnarray}  
Inserting the upper equation of (\ref{fundamental equation rederived}) into the left hand side of the above equation, and using Eq. (\ref{classical Hamiltonian measurement angular momentum}), one has, after an arrangement
\begin{eqnarray}
\partial_tS+g\big(x_1\partial_{y_1}S-y_1\partial_{x_1}S\big)\partial_{q_2}S-g\lambda^2\Big(x_1\frac{\partial_{y_1}\partial_{q_2}R}{R}\nonumber\\-y_1\frac{\partial_{x_1}\partial_{q_2}R}{R}\Big)+\frac{\lambda}{2\Omega}\Big(\partial_t\Omega-gy_1\partial_{x_1}(\Omega\partial_{q_2}S)\nonumber\\
+gx_1\partial_{y_1}(\Omega\partial_{q_2}S)+gx_1\partial_{q_2}(\Omega\partial_{y_1}S)-gy_1\partial_{q_2}(\Omega\partial_{x_1}S)\nonumber\\
-g\lambda(y_1\partial_{x_1}\partial_{q_2}\Omega-x_1\partial_{y_1}\partial_{q_2}\Omega)\Big)=0,
\label{ccc}
\end{eqnarray}
where $R=\sqrt{\Omega}$ and we have used the identity:
\begin{equation}
\frac{1}{4}\frac{\partial_{q_i}\Omega}{\Omega}\frac{\partial_{q_j}\Omega}{\Omega}=\frac{1}{2}\frac{\partial_{q_i}\partial_{q_j}\Omega}{\Omega}-\frac{\partial_{q_i}\partial_{q_j}R}{R}. 
\label{fluctuations decomposition}
\end{equation} 
Substituting Eq. (\ref{FPE angular momentum}), the last term of Eq. (\ref{ccc}) in the bracket vanishes to give 
\begin{eqnarray} 
\partial_tS+g\big(x_1\partial_{y_1}S-y_1\partial_{x_1}S\big)\partial_{q_2}S\hspace{20mm}\nonumber\\
-g\lambda^2\Big(x_1\frac{\partial_{y_1}\partial_{q_2}R}{R}-y_1\frac{\partial_{x_1}\partial_{q_2}R}{R}\Big)=0.
\label{HJM angular momentum}
\end{eqnarray} 

One thus has a pair of coupled equations (\ref{FPE angular momentum}) and (\ref{HJM angular momentum}) which are parameterized by $\lambda$. Recall that this pair of equations are valid in a microscopic time interval of length $\tau_{\xi}$ during which the magnitude of $\xi$ is constant while its sign is allowed to change randomly with equal probability. Averaging Eq. (\ref{FPE angular momentum}) for the cases $\pm\xi$, recalling that the sign of $\lambda$ is the same as that of $\xi$, one has, by the virtue of Eqs. (\ref{quantum phase symmetry}) and (\ref{God's fairness}), 
\begin{eqnarray}
\partial_t\Omega-gy_1\partial_{x_1}(\Omega\partial_{q_2}S_Q)+gx_1\partial_{y_1}(\Omega\partial_{q_2}S_Q)\hspace{10mm}\nonumber\\
+gx_1\partial_{q_2}(\Omega\partial_{y_1}S_Q)-gy_1\partial_{q_2}(\Omega\partial_{x_1}S_Q)=0.
\label{QCE for angular momentum measurement}
\end{eqnarray}
Similarly, averaging Eq. (\ref{HJM angular momentum}) over the cases $\pm\xi$, thus is also over $\pm\lambda$, will not change anything. We thus finally have a pair of Eqs. (\ref{HJM angular momentum}) and (\ref{QCE for angular momentum measurement}) which are now parameterized by $|\lambda|$ valid for a microscopic time interval of duration $\tau_{\xi}$ characterized by a constant $|\xi|$. 
 
Next, since $|\lambda|$ is non-vanishing, one can define the following complex-valued function:
\begin{equation}
\Psi\doteq \sqrt{\Omega}\exp\Big(i\frac{S}{|\lambda|}\Big). 
\label{general wave function} 
\end{equation}
Using $\Psi$ and recalling that $|\lambda|$ is constant during the microscopic time interval of interest with length $\tau_{\lambda}$,  the pair of Eqs. (\ref{HJM angular momentum}) and (\ref{QCE for angular momentum measurement}) can then be recast into the following compact form: 
\begin{equation}
i|\lambda|\partial_t\Psi=\frac{\lambda^2}{\hbar^2}\hat{H}_I\Psi.
\label{generalized Schroedinger equation angular momentum}
\end{equation}
Here $\hat{H}_I$ is a differential operator defined as 
\begin{equation}
{\hat H}_I\doteq g{\hat l}_{z_1}{\hat p}_2,
\label{Hamiltonian operator angular momentum}
\end{equation} 
where $\hat{p}_i\doteq-i\hbar\partial_{q_i}$, $i=1,2$ is the quantum mechanical momentum operator referring to the $i-$particle and ${\hat l}_{z_1}\doteq x_1{\hat p}_{y_1}-y_1{\hat p}_{x_1}$ is the $z-$part of the quantum mechanical angular momentum operator of the first particle; all are Hermitian. Recall that Eq. (\ref{generalized Schroedinger equation angular momentum}) is valid only for a microscopic time interval $[(k-1)\tau_{\xi},k\tau_{\xi})$ during which $|\xi|=|\xi_k|$ is constant. For finite time interval $t>\tau_{\xi}$, one must then apply Eq. (\ref{generalized Schroedinger equation angular momentum}) to each time intervals, each is parameterized by a random value of $|\xi_k|$, $k=1,2,\dots$, consecutively. 

Let us then consider a specific case when $|\lambda|$ is given by the reduced Planck constant $\hbar$, namely $\lambda=\pm\hbar$ with equal probability for all the time, so that the average of the deviation from infinitesimal stationary action distributed according to the exponential law of Eq. (\ref{exponential distribution of DISA}) is given by 
\begin{equation}
\hbar/2. 
\label{P}
\end{equation}
Moreover, let us assume that $P_{H}(\xi)$ is stationary in time and the fluctuations of $|\xi|$ around its average is sufficiently narrow so that $\Omega(q,|\xi|;t)$ and $S(q;t,|\xi|)$ can be approximated by the corresponding zeroth order terms in their Taylor expansion around the average of $|\xi|$, denoted respectively by $\rho_Q(q;t)$ and $S_Q(q;t)$. In this specific case, the zeroth order approximation of Eq. (\ref{generalized Schroedinger equation angular momentum}) then reads
\begin{eqnarray}
i\hbar\partial_t\Psi_Q(q;t)=\hat{H}_I\Psi_Q(q;t),\hspace{5mm}\nonumber\\
\Psi_Q(q;t)\doteq\sqrt{\rho_Q(q;t)}e^{\frac{i}{\hbar}S_Q(q;t)}.\hspace{0mm} 
\label{Schroedinger equation measurement of angular momentum}
\end{eqnarray}    
Unlike Eq. (\ref{generalized Schroedinger equation angular momentum}), Eq. (\ref{Schroedinger equation measurement of angular momentum}) is now deterministic parameterized by the reduced Planck constant $\hbar$. Moreover, from Eq. (\ref{Schroedinger equation measurement of angular momentum}), the Born's statistical interpretation of wave function is valid by construction
\begin{equation}
\rho_Q(q;t)=|\Psi_Q(q;t)|^2.
\label{Born's statistical interpretation}  
\end{equation} 
Equation (\ref{Schroedinger equation measurement of angular momentum}) together with Eq. (\ref{Hamiltonian operator angular momentum}) is just the Schr\"odinger equation for the von Neumann model of measurement of angular momentum of the first particle using the second particle as the apparatus. 

On the other hand, solving Eq. (\ref{generalized Schroedinger equation angular momentum}) for each time interval of length $\tau_{\xi}$ with a constant value of $|\xi|$, and inserting the modulus and phase of $\Psi$ into Eq. (\ref{velocity angular momentum}), we obtain the time evolution of the velocities of both of the particles which are randomly fluctuating due to the fluctuations of $\xi$. Recall again that Eq. (\ref{velocity angular momentum}) is valid in time interval of length $\tau_{\xi}$ in which the absolute value of $\xi$ is effectively constant while its sign fluctuates randomly with equal probability. It is then tempting to define an `effective' velocity as the average of the values of $\dot{q}$ at $\pm\xi$: 
\begin{equation}
\widetilde{\dot{q}}(|\xi|)\doteq\frac{\dot{q}(\xi)+\dot{q}(-\xi)}{2}. 
\label{effective velocity}
\end{equation}
When the actual velocities are given by Eq. (\ref{velocity angular momentum}), recalling that the sign of $\lambda$ is the same as that of $\xi$, one has, due to Eqs. (\ref{quantum phase symmetry}) and (\ref{God's fairness})
\begin{eqnarray}
\widetilde{\dot{x}_1}=-gy_1\partial_{q_2}S_Q,\hspace{2mm}\widetilde{\dot{y}_1}=gx_1\partial_{q_2}S_Q,\hspace{2mm}\nonumber\\
\widetilde{\dot{q}_2}=g\big(x_1\partial_{y_1}S_Q-y_1\partial_{x_1}S_Q\big),\hspace{5mm}
\label{Bohmian velocity angular momentum}
\end{eqnarray}
where we have counted only the zeroth order terms. Unlike Eq. (\ref{velocity angular momentum}), Eq. (\ref{Bohmian velocity angular momentum}) is now deterministic, due to the deterministic time evolution of $S_Q$ given by the Schr\"odinger equation of (\ref{Schroedinger equation measurement of angular momentum}). 

\subsection{A single measurement event, its ensemble and the Born's rule}

Let us now discuss the process of a single measurement event. From now on, we shall work with Eq. (\ref{Schroedinger equation measurement of angular momentum}) instead of with Eq. (\ref{generalized Schroedinger equation angular momentum}). To do this, let $\phi_l(q_1)$  denotes the eigenfunction of the angular momentum operator $\hat{l}_{z_1}$ belonging to an eigenvalue $\omega_l$: $\hat{l}_{z_1}\phi_l(q_1)=\omega_l\phi_l(q_1)$, $l=0,1,2,\dots$. $\{\phi_l\}$ thus makes a complete set of orthonormal functions. Then, ignoring the single particle Hamiltonians for impulsive interaction, the Schr\"odinger equation of (\ref{Schroedinger equation measurement of angular momentum}) has the following general solution: 
\begin{equation}
\Psi_Q(q_1,q_2;t)=\sum_lc_l\phi_l(q_1)\varphi(q_2-g\omega_lt), 
\label{entangled system-apparatus}
\end{equation}
where $\varphi(q_2)$ is the initial wave function of the apparatus (the second particle) which is assumed to be sufficiently localized in $q_2$, $\{c_l\}$ are complex numbers, and 
\begin{equation}
\phi(q_1)\doteq\sum_lc_l\phi_l(q_1),
\label{initial wave function of the system}
\end{equation} 
is the initial wave function of the system. $c_l$ is  thus the coefficient of expansion of the initial wave function of the system in term of the orthonormal set of the eigenfunctions of ${\hat l}_{z_1}$: 
\begin{equation}
c_l=\int dq_1\phi_l^*(q_1)\phi(q_1). 
\label{quantum amplitude}
\end{equation}
Initially, the total wave function is thus separable as \begin{equation}
\Psi_Q(q_1,q_2;0)=\Big(\sum_l c_l\phi_l(q_1)\Big)\varphi(q_2). 
\label{initial wave function} 
\end{equation}
It evolves into an inseparable (entangled) wave function of Eq. (\ref{entangled system-apparatus}) via the linear Schr\"odinger equation of (\ref{Schroedinger equation measurement of angular momentum}) with the interaction quantum Hamiltonian given by Eq. (\ref{Hamiltonian operator angular momentum}). One can then see in Eq. (\ref{entangled system-apparatus}) that for sufficiently large $g$, the set of wave functions 
\begin{equation}
\{\varphi_l(q_2;t_M)\doteq\varphi(q_2-g\omega_lt_M)\}, 
\end{equation}
are not overlapping for different $l$ and each is correlated to a distinct $\phi_l(q_1)$. 

Further, to have a physically and operationally smooth quantum-classical correspondence, one must let $q_2(t_M)$ has the same physical and operational status as the underlying classical mechanical system: namely, it must be regarded as the pointer of the measurement, the reading of our experiment. One may then {\it infer} that the `outcome' of a single measurement event corresponds to the packet $\varphi_l(q_2;t_M)$ whose support is actually entered by the apparatus particle. Namely, if $q_2(t_M)$ belongs to the spatially localized support of $\varphi_l(q_2;t_M)$, then we {\it operationally} admit that the result of measurement is given by $\omega_l$, the eigenvalue of $\hat{l}_{z_1}$ whose corresponding eigenfunction $\phi_l(q_1)$ is correlated with $\varphi_l(q_2;t_M)$. The probability that the measurement yields $\omega_l$ is thus equal to the frequency that $q_2(t_M)$ enters to the support of $\varphi_l(q_2;t_M)$ in a large (in principle infinite) number of identical experiments.  
  
It then remains to calculate the probability that $q_2(t_M)$ belongs to the support of $\varphi_l(q_2;t_M)$ given the initial wave function of the system $\phi(q_1)=\sum_lc_l\phi_l(q_1)$. To do this, first, since for sufficiently large value of $g$, $\{\varphi_l(q_2;t_M)\}$ in Eq. (\ref{entangled system-apparatus}) does not overlap for different values of $l$, then the joint-probability density that the first particle (system) is at $q_1$ and the second particle (apparatus) is at $q_2$ is, by the virtue of Eq. (\ref{Born's statistical interpretation}), decomposed into  
\begin{equation}
\rho_Q(q;t_M)=|\Psi_Q(q;t_M)|^2=\sum_l|c_l|^2|\phi_l(q_1)|^2|\varphi_l(q_2;t_M)|^2, 
\label{decomposition-approximation}
\end{equation}
namely the cross-terms are all vanishing. From the above equation, one can see that the joint-probability density that the first particle has coordinate $q_1$ and the second particle has coordinate $q_2$ inside the support of $\varphi_l(q_2;t_M)$ is given by  
\begin{equation}
|c_l|^2|\phi_l(q_1)|^2|\varphi_l(q_2;t_M)|^2. 
\label{joint-probability position-labelling}
\end{equation}
The probability density that the second particle is inside the support of the wave packet $\varphi_l(q_2;t_M)$ regardless of the position of the first and second particles is thus 
\begin{equation}
P_{\omega_l}=\int dq_1dq_2|c_l|^2|\phi_l(q_1)|^2|\varphi_l(q_2;t_M)|^2=|c_l|^2,
\label{Born's rule}
\end{equation}
which is just the Born's rule. 
  
\subsection{Discussion}

As noted at the beginning of the section, in reality the above model of measurement with one dimensional apparatus is oversimplified. Especially, the model excludes the irreversibility of the registration process which can only be done by realistic apparatus plus bath with large (macroscopic) degrees of freedom. See Ref. \cite{Theo} for an elaborated discussion of quantum measurement having realistic model of apparatus. 

Now, notice that as for the quantum Hamiltonian $\hat{H}_I$, the quantum mechanical angular momentum operator $\hat{l}_{z_1}$ appears formally when one works in Hilbert space by defining the wave function $\Psi$ as in Eq. (\ref{general wave function}), or its zeroth order term $\Psi_Q$, satisfying the linear Schr\"odinger equation of (\ref{Schroedinger equation measurement of angular momentum}). Hence, the Hermitian operator and wave function are artificial convenient mathematical tools for calculational purpose with no fundamental ontology. Moreover, the Hermitian angular momentum operator $\hat{l}_{z_1}$ emerges in the context of modeling a microscopic stochastic correction of the underlying classical mechanical model of measurement of angular momentum. The above results suggest that {\it not} all Hermitian operators are relevant for physics and conversely {\it not} all relevant and in principle observable physical quantities, such as position, time, mass etc., have to be represented by Hermitian operators. 

To have physically and operationally smooth correspondence with the underlying classical mechanical model of measurement discussed at the beginning of the section, we have kept regarding the position of the second particle as the pointer reading of the measurement. Namely the result of each single measurement is {\it inferred operationally} from the position of the pointer. We have shown however that unlike the classical mechanical case in which the result of measurement may take arbitrary continuous values, the set of possible values of the result of the measurement of angular momentum in the statistical model is discrete given by one of the eigenvalues of the angular momentum operator $\hat{l}_{z_1}$. The statistical model thus explicitly describes the physical and operational origin of the quantum discreteness of measurement results of angular momentum.  

Let us suppose that in a single measurement event $q_2(t_M)$ belongs to the support of $\varphi_l(q_2;t_M)$ so that we operationally infer that the measurement yields $\omega_l$. If the measurement is not destructive, then immediately after the first measurement, $q_2(t)$ will still belong to $\varphi_l(q_2;t)$, such that repeating the measurement will naturally yield the same value as the previous one, $\omega_l$.  Moreover, in this case, right after the measurement, $\Psi_{Q_l}\doteq\phi_l(q_1)\varphi_l(q_2;t)$ is the effective or relevant wave function of the whole system+apparatus. This is due to the fact that $q_2(t)$ is inside the support of $\varphi_l(q_2;t)$ which is not overlapping with $\varphi_{l'}(q_2;t)$, $l'\neq l$, and $q_2(t)$ can not pass through the nodes of the wave function. This situation is what is effectively regarded as the projection of the initial wave function $\phi(q_1)=\sum_l c_l\phi_l(q_1)$ of the system onto the corresponding eigenfunction $\phi_l(q_1)$ of the measurement result $\omega_l$, one of the eigenvalues of the angular momentum operator. We have thus an {\it effective} wave function collapse as one of the implication of the statistical model, rather than standing as an independent postulate as in standard quantum mechanics. Let us emphasize that during the process of measurement, the wave function of the whole system+apparatus satisfies the Schr\"odinger equation of (\ref{Schroedinger equation measurement of angular momentum}). There is thus {\it no real} wave function collapse. Hence, measurement is just a specific type of physical interaction following the same general law of dynamics and statistics. 

In the statistical model, the `actual' value of angular momentum prior to the measurement when the wave function is $\phi(q_1)=\sum_lc_l\phi_l(q_1)$ can take any continuous real numbers given by 
\begin{eqnarray}
l_{z_1}=x_1p_{y_1}-y_1p_{x_1}\hspace{50mm}\nonumber\\
=x_1\Big(\partial_{y_1}S_{Q_1}+\frac{\lambda}{2}\frac{\partial_{y_1}\Omega_{Q_1}}{\Omega_{Q_1}}\Big)-y_1\Big(\partial_{x_1}S_{Q_1}+\frac{\lambda}{2}\frac{\partial_{x_1}\Omega_{Q_1}}{\Omega_{Q_1}}\Big),
\label{actual angular momentum prior}
\end{eqnarray} 
where $\phi=\sqrt{\Omega_{Q_1}}\exp(iS_{Q_1}/\hbar)$, and we have used the upper equation in (\ref{fundamental equation rederived}). By contrast, in the above statistical model, each single measurement event will yield only discrete possible real numbers, one of the eigenvalues of angular momentum operator $\hat{l}_{z_1}$. Hence, one may conclude that the result of each single measurement does not in general reveal the `actual' value of angular momentum of the system prior to the measurement. 

Next, calculating the average of the angular momentum of the first particle prior to measurement over the distribution of the configuration, one gets
\begin{eqnarray}
\langle l_{z_1}\rangle\doteq
\int dq_1d\xi\Big(x_1\Big(\partial_{y_1}S_{Q_1}+\frac{\lambda}{2}\frac{\partial_{y_1}\Omega_{Q_1}}{\Omega_{Q_1}}\Big)\hspace{10mm}\nonumber\\
-y_1\Big(\partial_{x_1}S_{Q_1}+\frac{\lambda}{2}\frac{\partial_{x_1}\Omega_{Q_1}}{\Omega_{Q_1}}\Big)\Big)\Omega_{Q_1}\nonumber\\
=\int dq_1\Big(x_1\partial_{y_1}S_{Q_1}-y_1\partial_{x_1}S_{Q_1}\Big)\Omega_{Q_1}\hspace{10mm}\nonumber\\
=\int dq_1\phi^*(q_1)\hat{l}_{z_1}\phi(q_1), 
\label{average of actual am prior}
\end{eqnarray} 
where in the first equality we have used Eqs. (\ref{quantum phase symmetry}) and (\ref{God's fairness}), and taken into account the fact that the sign of $\lambda$ is the same as that of $\xi$. On the other hand, calculating the average of the results of measurement, one obtains
\begin{eqnarray}
\langle\hat{l}_{z_1}\rangle\doteq\sum_l\omega_lP_{\omega_l}=\sum_l\omega_l|c_l|^2\hspace{10mm}\nonumber\\
=\int dq_1\phi(q_1)^*\hat{l}_{z_1}\phi(q_1)=\langle l_{z_1}\rangle,
\end{eqnarray} 
where in the first equality we have used Eq. (\ref{Born's rule}), in the second equality we have used Eq. (\ref{initial wave function of the system}) and the last equality is just Eq. (\ref{average of actual am prior}). Hence, the average of measurement results in an ensemble of identical measurement is equal to the average of the actual value of angular momentum of the system prior to the measurement over the distribution of the configuration.   

Let us now ask: what is the actual value of the angular momentum of the first particle right after a measurement which yields $\omega_l$? Notice that in this case, the relevant wave function is given by $\phi_l$. Writing in polar form $\phi_l=\sqrt{\Omega_{Q_1}^l}\exp(iS_{Q_1}^l)$, one then has 
\begin{eqnarray}
l_{z_1}=x_1\Big(\partial_{y_1}S_{Q_1}^l+\frac{\lambda}{2}\frac{\partial_{y_1}\Omega_{Q_1}^l}{\Omega_{Q_1}^l}\Big)\hspace{10mm}\nonumber\\
-y_1\Big(\partial_{x_1}S_{Q_1}^l+\frac{\lambda}{2}\frac{\partial_{x_1}\Omega_{Q_1}^l}{\Omega_{Q_1}^l}\Big),
\label{actual angular momentum post}
\end{eqnarray}
Averaging its values at $\pm\xi$, the effective value reads 
\begin{eqnarray}
\widetilde{l_{z_1}}(|\xi|)\doteq \frac{l_{z_1}(\xi)+l_{z_1}(-\xi)}{2}=x_1\partial_{y_1}S_{Q_1}^l-y_1\partial_{x_1}S_{Q_1}^l\nonumber\\
=\frac{\mbox{Re}\{\phi_l^*\hat{l}_{z_1}\phi_l\}}{|\phi_l|^2}=\omega_l, 
\end{eqnarray}
where in the first equality we have made use Eqs. (\ref{quantum phase symmetry}) and (\ref{God's fairness}) and taken into account the fact that the sign of $\lambda$ is the same as that of $\xi$. Hence, when a measurement yields $\omega_l$, the value of the effective angular momentum $\widetilde{l_{z_1}}$ of the system right after the measurement is also given by $\omega_l$.    

Finally, let us mention without giving the technical details that the measurement of position and linear momentum can be done by exactly following all the steps for the measurement of angular momentum discussed above. One only needs to put $O_1=p_1$ and $O_1=q_1$ in Eq. (\ref{classical Hamiltonian measurement}) for the case of measurement of linear momentum and position, respectively. Repeating all the steps for the measurement of angular momentum, one will get a Schr\"odinger equation with the quantum Hamiltonian $\hat{H}_I=g\hat{O}_1\hat{p}_2$ where $\hat{O}_1=\hat{p}_1=-i\hbar\partial_{q_1}$ and $\hat{O}_1=\hat{q}_1=q_1$, respectively. All the qualitative and quantitative results for the measurement of angular momentum derived above then apply to the measurement of position and linear momentum. Note however that the spectrum of eigenvalues of $\hat{p}$ and $\hat{q}$ are continuous. The measurement of energy should be reduced to the measurement of position, linear and angular momentum.     
  
\section{Conclusion}   

We have developed a statistical model of microscopic stochastic deviation from classical mechanics based on a stochastic processes with a transition probability between two infinitesimally close spacetime points along a random path that is given by an exponential distribution of infinitesimal stationary action. We then applied the stochastic model to a classical mechanical model for the measurement of angular momentum. In the statistical model, the system always has a definite configuration all the time as in classical mechanics, following a randomly fluctuating continuous trajectory, regarded as the beable of the theory. On the other hand, we showed that the quantum mechanical Hermitian differential operator corresponding to the angular momentum arises formally together with the wave function as artificial convenient mathematical tools. In the model, the wave function is therefore neither physical nor complete. 

Reading the pointer of the corresponding classical system to operationally infer the results of the measurement, the model reproduces the prediction of quantum mechanics that each single measurement event yields randomly one of the eigenvalues of the Hermitian angular momentum operator with a probability given by the Born's rule. Moreover, during a single measurement event, the wave function of the system+apparatus evolves according to the Schr\"odinger equation so that there is no wave function collapse. We have thus a physically and operationally smooth correspondence between measurement in macroscopic and microscopic worlds.  

We have also shown that while the result of each single measurement event does not reveal the actual value of the angular momentum prior to measurement, its average in an ensemble of identical measurement is equal to the average of the actual value of the angular momentum prior to measurement over the distribution of the configuration. Moreover, we have shown that right after a single measurement, the effective value of angular momentum is equal to the result of measurement. 

Given the above results, it is then imperative to further ask, within the spirit of the reconstruction program, why the distribution of deviation from infinitesimal stationary action is given by the exponential law among the infinitude of possible distributions? Why Gaussian (say) will not work. It is then interesting to find a set of conceptually simple and physically transparent axioms which select uniquely the exponential law and to elaborate its relation with the characteristic traits of quantum mechanics. 

Finally, recall that the predictions of quantum mechanics are reproduced as the zeroth order approximation of the stochastic model for a specific choice of the free parameter of the transition probability so that the average deviation from the infinitesimal stationary action is given by $\hbar/2$. It is then interesting to go beyond the zeroth order term, and to elaborate the case when the average deviation from the infinitesimal stationary action is deviating slightly from $\hbar/2$. These cases might therefore provide precision tests against quantum mechanics. 
 
\begin{acknowledgments}   

\end{acknowledgments}  

\appendix

\section{Quantization of classical system of a single particle subjected to external potentials\label{app}}

To show the robustness of the model, we shall apply the statistical model to stochastically modify a classical system of a single particle subjected to external potentials so that the classical Hamiltonian takes the following form:
\begin{equation}
H(q,p)=\frac{g^{ij}(q)}{2}(p_i-a_i)(p_j-a_j)+V,  
\label{classical Hamiltonian}
\end{equation} 
where $a_i(q)$, $i=x,y,z$ and $V(q)$ are vector and scalar potentials respectively, the metric $g^{ij}(q)$ may depend on the position of the particle, and summation over repeated indices are assumed. This is a reproduction of the results reported in Ref. \cite{AgungSMQ4}. Note however that a couple of important assumptions put heuristically in Ref. \cite{AgungSMQ4}, that of Eqs. (\ref{quantum phase symmetry}) and (\ref{God's fairness}), are given physical argumentation in the present work. Below, we shall repeat all the steps we have taken to quantize the classical mechanical model for the measurement of angular momentum in the main text.      

Let us first consider a time interval of length $\tau_{\lambda}$ during which the absolute value of  $\lambda$ is effectively constant while its sign is allowed to fluctuate randomly together with the random fluctuations of the sign of $\xi$ in a time scale $dt$. Let us then divide it into a series of time intervals of length $\tau_{\xi}$, $[(k-1)\tau_{\xi},k\tau_{\xi})$, $k=1,2,\dots$ and attribute to each interval a random value of $\xi=\xi_k$ according to the probability distribution $P_{H_k}(\xi_k)=P_{H_k}(-\xi_k)$. Hence, in each interval, the magnitude of $\xi=\xi_k$ is constant while its sign is allowed to change randomly and the pair of equations in (\ref{fundamental equation rederived}) with fixed $|\xi_k|$ apply. 

Let us now consider a microscopic time interval $[(k-1)\tau_{\xi},k\tau_{\xi})$. Within this interval of time, using Eq. (\ref{classical Hamiltonian}) to express $\dot{q}$ in term of $p$ via the Hamilton equation $\dot{q}=\partial H/\partial p$, one has, by the virtue of the upper equation of (\ref{fundamental equation rederived})
\begin{equation}
\dot{q}^i(\xi)=g^{ij}\Big(\partial_{q_j}S(\xi)+\frac{\lambda(\xi)}{2}\frac{\partial_{q_j}\Omega(\xi)}{\Omega(\xi)}-a_j\Big). 
\label{classical velocity field HPF particle in potentials}
\end{equation}
Again, assuming the conservation of probability which is valid for the closed system we are considering, one obtains the following continuity equation:
\begin{eqnarray}
0=\partial_t\Omega+\partial_q\cdot(\dot{q}\Omega)\hspace{45mm}\nonumber\\
=\partial_t\Omega+\partial_{q_i}\Big(g^{ij}(\partial_{q_j}S-a_j)\Omega\Big)+\frac{\lambda}{2}\partial_{q_i}(g^{ij}\partial_{q_j}\Omega). 
\label{FPE particle in potentials}
\end{eqnarray} 

On the other hand, from Eq. (\ref{classical Hamiltonian}), $\theta(S)$ of Eq. (\ref{QC correspondence}) is given by 
\begin{equation}
\theta(S)=\partial_{q_i}g^{ij}(\partial_{q_j}S-a_j). 
\end{equation} 
The lower equation of (\ref{fundamental equation rederived}) thus becomes
\begin{eqnarray}
-H(q,p(\dot{q}))=\partial_tS+\frac{\lambda}{2}\frac{\partial_t\Omega}{\Omega}+\frac{\lambda}{2}\partial_{q_i}g^{ij}(\partial_{q_j}S-a_j).
\label{fundamental equation particle in potentials} 
\end{eqnarray}
Plugging the upper equation of (\ref{fundamental equation rederived}) into the left hand side of Eq. (\ref{fundamental equation particle in potentials}) and using Eq. (\ref{classical Hamiltonian}) one has, after an arrangement 
\begin{eqnarray}
\partial_tS+\frac{g^{ij}}{2}(\partial_{q_i}S-a_i)(\partial_{q_j}S-a_j)+V\hspace{30mm}\nonumber\\
-\frac{\lambda^2}{2}\Big(g^{ij}\frac{\partial_{q_i}\partial_{q_j}R}{R}+\partial_{q_i}g^{ij}\frac{\partial_{q_j}R}{R}\Big)\hspace{30mm}\nonumber\\
+\frac{\lambda}{2\Omega}\Big(\partial_t\Omega+\partial_{q_i}\Big(g^{ij}(\partial_{q_j}S-a_j)\Omega\Big)+\frac{\lambda}{2}\partial_{q_i}(g^{ij}\partial_{q_j}\Omega)\Big)=0,\nonumber\\
\label{HJM particle in potentials 0}
\end{eqnarray}
where $R\doteq\sqrt{\Omega}$ and we have again used the identity 
of Eq. (\ref{fluctuations decomposition}). Inserting Eq. (\ref{FPE particle in potentials}), the last line of Eq. (\ref{HJM particle in potentials 0}) vanishes to give
\begin{eqnarray}
\partial_tS+\frac{g^{ij}}{2}(\partial_{q_i}S-a_i)(\partial_{q_j}S-a_j)+V\nonumber\\
-\frac{\lambda^2}{2}\Big(g^{ij}\frac{\partial_{q_i}\partial_{q_j}R}{R}+\partial_{q_i}g^{ij}\frac{\partial_{q_j}R}{R}\Big)=0.
\label{HJM particle in potentials}
\end{eqnarray}

We have thus a pair of coupled equations (\ref{FPE particle in potentials}) and (\ref{HJM particle in potentials}) which are parameterized by $\lambda(\xi)$. Recall that the above pair of equations is valid in a microscopic time interval of length $\tau_{\xi}$ during which the magnitude of $\xi$ is constant while its sign is allowed to change randomly with equal probability. Moreover, recall also that the sign of $\lambda$ is always the same as the sign of $\xi$. Keeping this in mind, averaging Eq. (\ref{FPE particle in potentials}) for the cases $\pm\xi$, thus is also over $\pm\lambda$, one has, by the virtue of Eqs. (\ref{quantum phase symmetry}) and (\ref{God's fairness}), 
\begin{equation}
\partial_t\Omega+\partial_{q_i}\Big(g^{ij}(\partial_{q_j}S-a_j)\Omega\Big)=0. 
\label{QCE particle in potentials}
\end{equation}
Similarly, averaging Eq. (\ref{HJM particle in potentials}) for the cases $\pm\xi$ will not change anything. We have thus finally a pair of coupled equations (\ref{HJM particle in potentials}) and (\ref{QCE particle in potentials}) which are now parameterized by a constant $|\lambda|$, valid during a microscopic time interval of length $\tau_{\xi}$ characterized by a constant $|\xi|$. 
 
Using $\Psi$ defined in Eq. (\ref{general wave function}), and recalling the assumption that $|\lambda|$ is constant during the time interval of interest, the pair of Eqs. (\ref{HJM particle in potentials}) and (\ref{QCE particle in potentials}) can then be recast into the following modified Schr\"odinger equation: 
\begin{equation}
i|\lambda|\partial_t\Psi=\frac{1}{2}(-i|\lambda|\partial_{q_i}-a_i)g^{ij}(q)(-i|\lambda|\partial_{q_j}-a_j)\Psi+V\Psi. 
\label{generalized Schroedinger equation particle in potentials}
\end{equation} 
Notice that the above equation is valid only for a microscopic time interval $[(n-1)\tau_{\xi},n\tau_{\xi})$ during which the magnitude of $\xi=\xi_n$ is constant. For finite time interval $t>\tau_{\xi}$, one must then apply Eq. (\ref{generalized Schroedinger equation particle in potentials}) consecutively to each time intervals of length $\tau_{\xi}$ with different random values of $|\xi_n|$, $n=1,2,3,\dots$. 

Let us again consider a specific case when $|\lambda|=\hbar$ so that the average of the deviation from infinitesimal stationary action distributed according to the exponential law of Eq. (\ref{exponential distribution of DISA}) is given by $\hbar/2$. The zeroth order approximation of Eq. (\ref{generalized Schroedinger equation particle in potentials}) then reads  
\begin{eqnarray}
i\hbar\partial_t\Psi_Q(q;t)=\hat{H}\Psi_Q(q;t),
\label{Schroedinger equation particle in potentials} 
\end{eqnarray}
where $\Psi_Q$ is defined as in Eq. (\ref{Schroedinger equation measurement of angular momentum}) and $\hat{H}$ is the quantum Hamiltonian given by 
\begin{equation}
\hat{H}=\frac{1}{2}(\hat{p}_i-a_i)g^{ij}(q)(\hat{p}_j-a_j)+V. 
\label{quantum Hamiltonian particle in potentials}
\end{equation}
Unlike Eq. (\ref{generalized Schroedinger equation particle in potentials}), Eq. (\ref{Schroedinger equation particle in potentials}) is now deterministic parameterized by $\hbar$. One can also see that unlike canonical quantization which, for the general type of $g^{ij}(q)$, suffers from the problem of operator ordering ambiguity, the resulting quantum Hamiltonian is unique in which $g^{ij}(q)$ is sandwiched by $\hat{p}-a$.     

Solving the modified Schr\"odinger equation of (\ref{generalized Schroedinger equation particle in potentials}) for each time interval of length $\tau_{\xi}$ with a fixed value of $|\xi|$, and inserting the modulus and phase of $\Psi$ into Eq. (\ref{classical velocity field HPF particle in potentials}), one obtains the stochastic evolution of the velocity of the particle as $\Psi$ evolves with time. In this case, the effective velocity defined in Eq. (\ref{effective velocity}) reads
\begin{equation}
\widetilde{\dot{q}}^i(|\xi|)\doteq\frac{\dot{q}^i(\xi)+\dot{q}^i(-\xi)}{2}=g^{ij}\Big(\partial_{q_j}S(|\xi|)-a_j\Big), 
\label{effective classical velocity field HPF particle in potentials}
\end{equation}    
the zeroth order approximation of which gives 
\begin{equation}
\widetilde{\dot{q}}^i=g^{ij}\Big(\partial_{q_j}S_Q-a_j\Big),
\end{equation}
which, unlike Eq. (\ref{classical velocity field HPF particle in potentials}), is now deterministic, due to the deterministic time evolution of $S_Q$ given by the Schr\"odinger equation of (\ref{Schroedinger equation particle in potentials}).

\end{document}